\documentclass[11pt]{JHEP3}

\usepackage[latin1]{inputenc}
\usepackage{nicefrac}
\usepackage{amsfonts}
\usepackage{amsmath}
\usepackage{amssymb}
\usepackage{graphicx}
\usepackage{epsfig, multicol}
\usepackage{appendix}
\usepackage{cite}

\providecommand{\U}[1]{\protect\rule{.1in}{.1in}}
\numberwithin{equation}{section}
\title{On S-Matrix factorization of the Landau-Lifshitz model}
\author{A. Melikyan, A. Pinzul, V. O. Rivelles, G. Weber\thanks{
melikyan,apinzul,rivelles,weber@fma.if.usp.br} \\
\\
\em{Instituto de Física}\\
\em{Universidade de São Paulo}\\
\em{C. Postal 05315-970, São Paulo, SP, Brazil}\\
}
\date{}
\abstract{
We consider the three-particle scattering S-matrix for the Landau-Lifshitz
model by directly computing the set of the Feynman diagrams up to the second
order. We show, following the analogous computations for the non-linear Schrödinger model \cite{Thacker:1974kv, Thacker:1976vp}, that the three-particle S-matrix is factorizable in the first non-trivial order.}
\preprint{}
\begin{document}
\section{Introduction}

The Landau-Lifshitz (LL) model describes the dynamics of the classical
Heisenberg spin chain, and has appeared in the past few years in both gauge
and string theories as one of the first indications of the underlying
integrable structure \cite{Kazakov:2004qf,Gromov:2006dh,Gromov:2006cq,Arutyunov:2004vx,Staudacher:2004tk,Arutyunov:2006iu,Janik:2006dc,Beisert:2006zy,Schafer-Nameki:2004ik,Beisert:2003ys,Bellucci:2006bv,Astolfi:2008yw,Harmark:2008gm}
of the AdS/CFT correspondence (for a review see \cite{Tseytlin:2004xa,
Tseytlin:2003ii, Minahan:2006sk, Zarembo:2004hp}). On the string theory side
the LL model emerges in the $R\times S^{3}$ subsector in the limit of large
angular momentum \cite{Kruczenski:2003gt, Kruczenski:2004kw, Kazakov:2004qf,
Roiban:2006yc, Stefanski:2007dp, Tirziu:2006ve, Fradkin:1991nr}.
Alternatively, the LL model can be derived from the Faddeev-Reshetikhin (FR)
model in the low-energy limit \cite{Faddeev:1985qu, Klose:2006dd}. The
latter is aimed to resolve the difficulties of the quantization procedure 
\cite{Faddeev:1982rn,Faddeev:1987ph,Mikhailov:2005sy, Mikhailov:2007eg,
Maillet:1985ek,Duncan:1989vg, Das:2004hy,Das:2005hp, Das:2007tb} inherent to
all sigma models. More recently, the LL model also emerged in the $SU(2)\times SU(2)$ subsector of the strings on $AdS_{4}\times\mathbb{C}P^{3} 
$ \cite{Grignani:2008is} in the context of the newly proposed duality
between $\mathcal{N}=6$ superconformal Chern-Simons and the $AdS_{4}\times 
\mathbb{C}P^{3}$ string theories in the t'Hooft limit \cite{Aharony:2008ug}.
The integrable properties of the LL model have been extensively discussed
from various points of view. The classical integrability was established for
the isotropic case in \cite{Takhtajan:1977rv}, and for the general
anisotropic case in \cite{Sklyanin:1979ll,Faddeev:1987ph}, and the
quasi-classical analysis was performed in \cite{Jevicki:1978yv}. The
classical equivalence between the LL and the non-linear Schrödinger (NLS)
models was shown in \cite{Zakharov:1979jc} by constructing a gauge
transformation between the flat currents of the corresponding models.

At the quantum level, on the other hand, the integrable properties of the LL
and NLS models are quite different. In \cite{Sklyanin:1988s1} Sklyanin has
considered in detail the quantum inverse scattering method for the LL model
(the $su(1,1)$ case), and pointed out several subtleties of the quantization
procedure, absent in the classically equivalent NLS model. One of the
surprising difficulties is that the standard methods to obtain the
Yang-Baxter relation from its classical counterpart fail for the anisotropic
LL model, and the algebra of observables, as well as the transfer matrix
have to be modified by hand in order to construct the quantum R-matrix. More
important implication of the Sklyanin's analysis is the problem of
constructing the local integrals of motion in the quantum theory, underlying
the quantum integrability of the model. In fact, the problems already arise
in constructing the local quantum Hamiltonian for the two-particle sector,
and to the best of our knowledge, the higher order charges have not been
constructed so far. The difficulties arise as a result of the ill-defined
operator product at the same point\footnote{For more details see the upcoming paper \cite{MelPin}} and can, in
principle, be resolved by the standard procedure of putting the model on the
lattice \cite{Izergin:1982ry, Izergin:1981mc}. Although this can be done for
both FR and LL models \cite{Faddeev:1985qu, Korepin:1997bk, Kulish:1981bi},
and the corresponding spectra can be found with the standard Bethe Ansatz
technique, the full $AdS_{5}\times S^{5}$ string model is too complicated,
and in general putting a continuous theory on the lattice using the standard
methods leads to a non-local quantum Hamiltonian \cite{Izergin:1982ry,
Gaudin:1983}. Thus, one has to deal with the quantization of the continuous
models directly, and the LL model, despite its simplicity, is a particularly
interesting representative example of the difficulties associated with the continuous quantum inverse scattering
method.

The LL and FR models, as well as a number of other models, arising in
various limits of the full $AdS_{5}\times S^{5}$ string, have recently been
considered as two-dimensional field theories, and the corresponding
S-matrices have been obtained by means of perturbative calculations \cite{Klose:2006dd,Klose:2006zd, Klose:2007wq, Klose:2007rz, Puletti:2007hq}. The
idea of this method and its relation to the inverse scattering method is
nicely reviewed for the NLS model in \cite{Thacker:1980ei}. This method
makes use of the integrability, which is believed to hold at the quantum
level, and, in particular, the S-matrix factorization property \cite{Zamolodchikov:1978xm, Kulish:1975ba, Cherednik:1980ey, Dorey:1996gd,
Parke:1980ki, Karowski:1978vz, Karowski:1977tv, Karowski:1978eg}, which
renders the N-particle scattering S-matrix to be a product of the
two-particle S-matrices. The direct verification of the factorization
property, by calculating the set of Feynman diagrams, has only been possible to
carry out for the NLS model in \cite{Thacker:1974kv, Thacker:1976vp}. More
recently, the three-particle S-matrix factorization property was
demonstrated for the $AdS_{5}\times S^{5}$ string in the near flat space
limit at one-loop order \cite{Puletti:2007hq}.

In this article we consider the three-particle scattering process for the LL
model and, by computing the necessary Feynman diagrams up to the second
order, show that it is factorizable in the first non-trivial order. There
are several conceptual features that make the S-matrix calculations for the
LL model a non-trivial task. First, as we noted earlier, the LL model is the
low energy limit for the FR model, and as was shown in \cite{Das:2007tb},
the Hamiltonian for the FR model is not diagonalizable in the class
of the standard representation for the two-particle sector: 
\begin{equation}
|p_{1}p_{2}\rangle=\underset{x_{1}\neq x_{2}}{\iint}dx_{1}dx_{2}\left[
e^{p_{1}x_{1}+p_{2}x_{2}}+\left( S\right) e^{p_{1}x_{2}+p_{2}x_{1}}\right]
\phi^{+}(x_{1})\phi^{+}(x_{2})|0\rangle  \label{2pwf}
\end{equation}

This representation is at the heart of the Bethe Ansatz and has the clear
interpretation for the first and second terms as the incoming and outgoing
waves respectively, and where the S-matrix $S(p_{1},p_{2})=e^{i\Delta
(p_{1},p_{2})}$ is a phase-shift due to the scattering (we assume that $p_{1}>p_{2}$). Proceeding in a similar manner to \cite{Das:2007tb} it is not
difficult to show that the Hamiltonian for the LL model cannot be
diagonalized for the class of functions (\ref{2pwf} ), although the exact
two-particle scattering S-matrix for the FR model can be obtained using the
perturbative calculations \cite{Klose:2006dd}. Alternatively, this can also
be seen directly  via the quantum inverse scattering method \cite{Sklyanin:1988s1}, where the two-particle wave function is shown to acquire
an additional term to (\ref{2pwf}). One of the reasons why this happens is
the extremely singular nature of the LL interaction \cite{Sklyanin:1988s1}.
In fact, it is not difficult to see that the interaction in the quantum
mechanical picture corresponds to the second derivative of the
delta-function, and dealing with it is not an easy task (see for example \cite{Albeverio:1988}). In order to write the additional term to (\ref{2pwf}), one has to
introduce a new creation operator, which for the two-particle state would
correspond to a bound state. It is not difficult to see that the number of
such terms increases with the number of particles in the scattering process.
For example, in the three-particle scattering process there will be two
additional contributions, corresponding to three and two-particle clusters.

More importantly, the form (\ref{2pwf}) suggests that the particles created
by the fields $\phi ^{+}(x)|0\rangle $ are those corresponding to the Bethe
particles, and all the consequences of the integrability, such as the
S-matrix factorization, will hold as the consequence of the general argument
of \cite{Zamolodchikov:1978xm,Kulish:1975ba,Parke:1980ki}. This is indeed
the case for a number of models, e.g. the NLS model, where (\ref{2pwf})
is indeed the two-particle state, and the particles $\phi ^{+}(x)|0\rangle $
are the Bethe particles, for which one can obtain the S-matrix either using
the perturbative diagrammatic calculations, or via direct diagonalization of
the Hamiltonian. However, this \emph{a priori} is not the case in general,
and the particles created by the fields $\phi ^{+}(x)$ may not correspond to
the Bethe particles. This is the case for the LL model, namely, the Bethe
particles do not coincide with $\phi ^{+}(x)|0\rangle $. Let us emphasize
that considering the scattering process and stating the S-matrix
factorization makes sense only in terms of the Bethe particles. The only
reliable method for diagonalizing all the local conserved quantities
simultaneously, including the Hamiltonian, and, as a consequence,
constructing the eigenstates corresponding to the Bethe particles, is the
quantum inverse scattering method. But as we mentioned above, this
construction fails in the LL model, and the local conserved quantities
cannot be derived from the trace identities due to the singular expressions.
Thus, checking the quantum integrability, e.g., the factorization of the
S-matrix, as well as establishing a connection with the construction in the
quantum inverse scattering method by using the standard field theoretic
methods is an important task that should prove useful when considering more
complicated systems than the LL model. In this paper and \cite{MelPin} we
make the first steps towards this program, and as the first result we show here
the factorization in the lowest order.

There are other subtleties, making factorization for the LL model a
non-trivial feature. The unconstrained two dimensional field theory on the
world-sheet corresponding to the LL model contains infinite number of
vertices. This is in contrast with the simpler NLS model, where there is
simply one vertex to deal with, and, thus, making the higher order
calculations much simpler. In the LL model one has to correspondingly
consider new types of vertices as the order of calculations increases. Let
us recall, that to calculate the two-particle scattering S-matrix one only
needs to keep the terms up to the quartic order in the Lagrangian. The role of higher order
terms is to preserve the integrability at the quantum level. Namely, for the
three-particle scattering S-matrix one has to consider the vertices of
higher (up to $\phi ^{6}$) order. The quantum integrability manifests here
in the S-matrix factorization for the three-particle scattering. Although
the combinatorial analysis due to the infinite number of vertices makes
practically impossible to carry out the calculations to all orders, we are able to show in
the first non-trivial order that the higher order vertices are exactly of
the form needed for the S-matrix to be factorizable. The mechanism behind
factorization is not straightforward, and we show that the higher order
terms cancel unwanted lower order terms to guarantee the factorization.
Moreover, we show explicitly that there is no process corresponding to
particle annihilation or creation, and the set of momenta before and after
the scattering is the same, in agreement with our understanding of
integrability.

Our paper is organized as follows. In Section \textbf{\ref{LL2part}}, we
give a brief account of the perturbative S-matrix calculations for the
Landau-Lifshitz model in the two-particle sector. In Section \textbf{\ref{LL3part}}, we present our analysis of the three-particle scattering for the
LL model and show the factorization in the first non-trivial order. In
section \textbf{\ref{conclusion}}, we give a brief summary of our results.
In the \textbf{Appendix} we collect the set of the Feynman graphs used in
our calculations in order to avoid cluttering the main text.

\section{Two-particle scattering S-matrix}

\label{LL2part}

In this section we set up the notation and present the necessary formulas
for the subsequent section on the three-particles scattering. First we
briefly review the necessary facts about the LL model, and then explain the
two-particle perturbative S-matrix calculations, referring the reader to 
\cite{Klose:2006dd} for complete details of the calculations. Let also us
note, that the two-particle scattering diagrammatic calculations for the LL
model are quite similar to those of the non-linear Schrödinger model \cite{Thacker:1980ei}. Indeed, the only essential difference is the presence of
the derivatives in the interaction terms. In the context of the perturbative
calculations this results in a slightly more complex combinatorial analysis in
the LL model. As we will see below, the essential differences arise when
considering the three-particle scattering amplitude.

\subsection{Landau-Lifshitz model: preliminary facts}


The Landau-Lifshitz model is a non-relativistic sigma model which describes
the continuous Heisenberg ferromagnet, and arises in the $R\times S^{3}$
subsector of the $AdS_{5}\times S^{5}$ strings in the limit of the large
angular momentum. The equations of motion for the isotropic LL model, on
which we will focus in this paper, in terms of the spin variables $\mathbf{S( }x\mathbf{)}=\left( S^{a}(x);a=1,2,3\right) ,$ have the form: 
\begin{equation}
\partial_{t}\mathbf{S}=\mathbf{S\times}\partial_{x}^{2}\mathbf{S}
\label{LL0}
\end{equation}
where the fields $\mathbf{S(}x\mathbf{)}$ take values on $S^2$: 
\begin{equation}
\mathbf{S}^{2}=1  \label{LL0aa}
\end{equation}
The equation (\ref{LL0}) can be obtained from the Hamiltonian: 
\begin{equation}
H_{LL}=\frac{1}{4}\int dx(\partial_{x}\mathbf{S})^{2}  \label{LL0a}
\end{equation}
with the Poisson structure: 
\begin{equation}
\{S^{a}(x),S^{b}(y)\}=-\varepsilon^{abc}S^{c}\delta(x-y)  \label{LL0b}
\end{equation}
In the most general formulation, the equations of motion for the anisotropic LL\
model are modified by an anisotropy matrix $\mathbf{J}=\text{diag}
(J_1,J_2,J_3),$ and take the form: 
\begin{equation}
\partial_{t}\mathbf{S}=\mathbf{S\times}\partial_{x}^{2}\mathbf{S+S\times JS}
\label{LL0c}
\end{equation}
As explained in \cite{Sklyanin:1988s1} the isotropic and anisotropic cases
should be considered separately when quantizing the theory. In the former
case, the standard inverse scattering procedure goes through without any
changes, and the Yang-Baxter equation is satisfied with the appropriate
choice of the $R$-matrix, while in the later case the operators, and
accordingly the operator algebra (\ref{LL0b}) have to be modified by hand
for the Yang-Baxter relation to have a solution. In this paper we will
consider only the isotropic case $\mathbf{J=0}$, and the more general case
is currently under investigation.

The corresponding action can be written in a manifestly covariant form as
follows:
\begin{equation}
S=\int d^{2}x\left[ C_{t}(\mathbf{S})-\frac{1}{4}(\partial_{x}\mathbf{S}
)^{2} \right]  \label{LL1}
\end{equation}
where $C_{t}(\mathbf{S})$ is the Wess-Zumino term:

\begin{equation}
C_{t}(\mathbf{S})=-\frac{1}{2}\int_{0}^{1}d\xi\;\epsilon_{ijk}S_{i}
\partial_{\xi}S_{j}\partial_{t}S_{k}  \label{LL2}
\end{equation}
The boundary conditions for the $\mathbf{S}(t,x;\xi)$ field have the form:

\begin{equation}
\left\{ 
\begin{array}{ccc}
\mathbf{S}(t,x;\xi=1) & = & \mathbf{S}_{0} \\ 
\mathbf{S}(t,x;\xi=0) & = & \mathbf{S}(t,x)
\end{array}
\right.  \label{LL2a}
\end{equation}
where $\mathbf{S}_{0}\ $is a constant vector.

Following \cite{Klose:2006dd}, it is convenient to resolve the constraint (\ref{LL0aa}) and write the action (\ref{LL1}) as an unconstrained $(1+1)$-dimensional quantum field theory. This can be achieved by a change of variables, which get rid of the non-linearities in the kinetic term, as follows 
\begin{equation}
\varphi=\frac{S_{1}+iS_{2}}{\sqrt{2+2S_{3}}}\quad,\quad
S_{3}=1-2|\varphi|^{2}  \label{LL5}
\end{equation}
and which turns the Wess-Zumino term (\ref{LL2a}) into a but non-covariant
form. The resulting action following from (\ref{LL1}) and (\ref{LL5}) can be
written as follows: 
\begin{equation}
S = \int d^{2}x \: \left\{ \frac{i}{2} (\varphi^{*} \partial_{t} \varphi-
\partial_{t} \varphi^{*} \varphi) - |\partial_{x} \varphi|^{2} -\frac{1}{4} 
\frac{2 -|\varphi|^{2}}{1 - |\varphi|^{2}} \left[ (\varphi
^{*}\partial_{x}\varphi)^{2} +(\partial_{x} \varphi^{*} \varphi)^{2} \right]
-\frac{1}{2} \frac{|\varphi|^{4}|\partial_{x} \varphi|^{2}}{1 - |\varphi|^{2}}\right\}  \label{LL6}
\end{equation}

The non-relativistic character of the LL model leads to significant
simplifications in the quantization procedure. In particular, it makes
diagrammatic calculations quite easy to deal with. This is mostly due to the
property of the propagator in the LL model to be a retarded one. Let us also
note, that in relativistic theories, such as the massive Thirring model, to
have similar simplifications in diagrammatic calculations, one can choose
the false vacuum by hand, and, therefore, making the Feynman propagators to
be retarded.

\begin{figure}[tbp]
\centering
\includegraphics[scale=0.55]{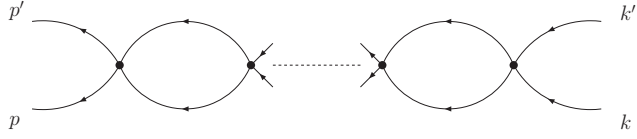}
\caption{Typical bubble diagram for the two-particle S-matrix.}
\label{bubbles}
\end{figure}

Using the field decomposition

\begin{equation}
\varphi (x)=\int \frac{dp}{2\pi }e^{-ip\cdot x}a_{p}\quad ,\quad \varphi
^{\ast }(x)=\int \frac{dp}{2\pi }e^{ip\cdot x}a_{p}^{\dagger }  \label{LL7}
\end{equation}
where we use the notation $p\cdot x=p^{2}x^{0}-px,$ and defining the vacuum of the
theory by $\varphi (x)|0\rangle =0$, the annihilation and
creation operators $a_{p}$ and $a_{p}^{\dagger }$ satisfy the standard
commutation relation:

\begin{equation}
\lbrack a_{p^{\prime}},a_{p}^{\dagger}]=2\pi\delta(p-p^{\prime}).
\label{LL8}
\end{equation}
One then finds the propagator to be of the form \cite{Klose:2006dd}:

\begin{align}
D(t,x) & =\int\frac{d^{2}q}{\left( 2\pi\right) ^{2}}\;\frac{ie^{-iq\cdot x}}{q^{0}-q^{2}+i\epsilon}  \notag \\
&  \label{LL9a} \\
& =\theta(t)\sqrt{\frac{\pi}{it}}e^{\frac{ix^{2}}{4t}}  \notag
\end{align}
The fact that the propagator is purely retarded leads to the following
result: the two particle scattering S-matrix is given by the sum of bubble
diagrams, as in Fig. \ref{bubbles}.

The diagram above is exactly the one that appears in the NLS model. For the
LL model one has to properly take into account the presence of the
derivatives in the interaction terms, which results in four different types
of vertices, depending on the placement of the derivatives on the internal
or external lines. Nevertheless, the calculations are still easy to carry
out.

\subsection{Diagrammatic calculations}


Since the only contributions to the two-particle scattering amplitude are the
diagrams of the type Fig. \ref{bubbles}, to calculate the two-particle
S-matrix it is enough to truncate the complicated action (\ref{LL6}) up to
the fourth order in the fields:

\begin{equation}
S=\int d^{2}x\;\frac{i}{2}(\varphi^{\ast}\partial_{t}\varphi-\partial
_{t}\varphi^{\ast}\varphi)-|\partial_{x}\varphi|^{2}-\frac{g}{2}[(\varphi^{\ast}\partial_{x}\varphi)^{2}+(\partial_{x}\varphi^{\ast}
\varphi)^{2}]+O(\varphi^{6})  \label{LL11}
\end{equation}
We introduced in the expansion (\ref{LL11}) the formal parameter $g$ to keep
track of the perturbative order, and it should be set to one at the end of the calculations. Let us
note here, that this is very important for the three-particle scattering
amplitude factorization, where several such formal parameters will be
introduced to take care of order counting for different types of vertices
that arise in the LL model when expanding the action (\ref{LL6}) up to the
sixth order in the fields. A natural question arises whether setting the
parameters to one at the end is consistent with the renormalization
properties of the LL model. Although we ignore here all the divergences
associated with the loop calculations, this is an interesting question and
is currently under investigation.

We collect here the necessary expressions, referring the reader to \cite{Klose:2006dd} for complete details. Taking without any loss of generality
the ordering $p>p^{\prime }$ for the scattering to take place, the
two-particle S-matrix is determined from the relation

\begin{equation}
\langle kk^{\prime}|\hat{S}|pp^{\prime}\rangle=S(p,p^{\prime})\delta_{+}^{(2)}(p,p^{\prime};k,k^{\prime})  \label{LL12}
\end{equation}
where:

\begin{equation}
\delta _{+}^{(2)}(p,p^{\prime };k,k^{\prime })=(2\pi )^{2}\left[ \delta
(p-k)\delta (p^{\prime }-k^{\prime })+\delta (p-k^{\prime })\delta
(p^{\prime }-k)\right]  \label{LL13}
\end{equation}
The scattering amplitude given by: 
\begin{align}
\langle kk^{\prime }|\hat{S}|pp^{\prime }\rangle & =\langle kk^{\prime
}|Te^{-i\int H_{int}dt}|pp^{\prime }\rangle  \notag  \label{LL13a} \\
& =\langle kk^{\prime }|pp^{\prime }\rangle -i\langle kk^{\prime }|T\int
H_{int}dt|pp^{\prime }\rangle -\frac{1}{2}\langle kk^{\prime }|T\left( \int
H_{int}dt\right) ^{2}|pp^{\prime }\rangle +\cdots
\end{align}
is easily computed with $H_{int}=\frac{g}{2}\int dx\;\left[ (\varphi ^{\ast
}\partial _{x}\varphi )^{2}+(\partial _{x}\varphi ^{\ast }\varphi )^{2}\right] $ in each order. The non-scattering part and the tree level parts
are easily computed and have the forms:

\begin{equation}
\langle kk^{\prime}|pp^{\prime}\rangle=\delta_{+}^{(2)}(p,p^{\prime
};k,k^{\prime})  \label{LL14}
\end{equation}

\begin{equation}
-i\langle kk^{\prime }|T\int H_{int}dt|pp^{\prime }\rangle =2ig\frac{pp^{\prime }}{p-p^{\prime }}\delta _{+}^{(2)}(p,p^{\prime };k,k^{\prime })
\label{LL15}
\end{equation}
To avoid the combinatorial analysis of \cite{Klose:2006dd}, associated with
placement of derivatives on internal and external lines, we will compute the
full bubble diagram in the following manner. Using the complete interaction
vertex, which in the momentum representation has the form: 
\begin{equation}
V(k,k^{\prime };p,p^{\prime })=2ig\left[ pp^{\prime }+kk^{\prime }\right]
\delta (p^{0}+p^{\prime 0}-k^{0}-k^{\prime 0})\delta (p+p^{\prime
}-k-k^{\prime }),  \label{LL16}
\end{equation}
we compute the complete one-loop diagram, while keeping the external line
corresponding to $k$ and $k^{\prime }$ off-shell, and putting the momenta $p$
and $p^{\prime }$ on-shell. The resulting expression has the following form: 
\begin{equation}
\langle kk^{\prime }|\hat{S}|pp^{\prime }\rangle \big|_{g^{2}}=\left[
2ig\left( pp^{\prime }+kk^{\prime }\right) (2\pi )^{2}\delta ^{(2)}(p+p^{\prime
}-k-k^{\prime })\right] ig\frac{2pp^{\prime }}{p-p^{\prime }}(pp^{\prime
}+kk^{\prime })  \label{LL16a}
\end{equation}
This already includes the sum of all one-loop diagrams with all possible
placement of derivatives on internal and external lines, with the correct
combinatorial factors. Noting that the term inside the square brackets in (\ref{LL16a}) coincides with the expression for the interaction vertex (\ref{LL16}), one can regard the one-loop scattering amplitude as the
interaction vertex in momentum representation, with the momenta $p$ and $p^{\prime }$ on-shell, multiplied by some function of this pair of momenta
and the expansion parameter. So that the $n$-loop scattering amplitude
corresponds to the product of $n$ of these modified vertices.

\begin{equation}
\langle kk^{\prime }|\hat{S}|pp^{\prime }\rangle \big|_{g^{n+1}}=\left( ig\frac{pp^{\prime }}{p-p^{\prime }}\right) ^{n}2ig(2\pi )^{2}\delta
^{(2)}(p+p^{\prime }-k-k^{\prime })(pp^{\prime }+kk^{\prime })  \label{LL17}
\end{equation}
If we also put $k$ and $k^{\prime }$ on-shell, we find that:

\begin{equation}
\langle kk^{\prime }|\hat{S}|pp^{\prime }\rangle \big|_{g^{n+1}}=2\left( ig\frac{pp^{\prime }}{p-p^{\prime }}\right) ^{n+1}\delta
_{+}^{(2)}(p,p^{\prime };k,k^{\prime }),  \label{LL19a}
\end{equation}
where we have used the relation
\begin{equation}
(2\pi)^{2}\delta^{(2)}(p+p^{\prime}-k-k^{\prime})=\frac{1}{2(p-p^{\prime}%
)}\delta_{+}^{(2)}(p,p^{\prime};k,k^{\prime})\label{rel1}
\end{equation}
Thus, two-particle scattering S-matrix has the form \cite{Klose:2006dd}.

\begin{equation}
S(p,p^{\prime })=1+2\sum_{n=1}^{\infty }(ig)^{n}\left( \frac{pp^{\prime }}{
p-p^{\prime }}\right) ^{n}=\frac{\frac{1}{p}-\frac{1}{p^{\prime }}-ig}{\frac{
1}{p}-\frac{1}{p^{\prime }}+ig}\overset{g\rightarrow 1}{\longrightarrow } 
\frac{\frac{1}{p}-\frac{1}{p^{\prime }}-i}{\frac{1}{p}-\frac{1}{p^{\prime }}
+i}  \label{LL20}
\end{equation}

\section{Three-particle scattering S-matrix}

\label{LL3part}

The direct verification of the S-matrix factorization using the perturbative
calculations is quite difficult to carry out, due to the complicated
diagrammatic analysis. The only model for which the factorization property
was manifestly verified is the non-linear Schrödinger model. We refer to the
details of the diagrammatic calculations for the case of the NLS model to 
\cite{Thacker:1974kv,Thacker:1976vp}. Before presenting the concrete
calculations for the LL model let us note that there is a significant
difference between the NLS and LL models, and the latter case is
considerably more complex. Indeed, the Hamiltonian for the NLS model has the
form: 
\begin{equation}
H=\int dx\;|\partial _{x}\phi (x)|^{2}+c|\phi (x)|^{4}  \label{LL3:0}
\end{equation}
Thus, while calculating the three-particle scattering amplitude, one has to
deal with only one type of vertex. The same is true when considering the
general $N$-particle scattering process. In the LL model the situation is
different and to consider the three-particle scattering amplitude one has
to truncate the full action (\ref{LL6}) up to the sixth order in fields. The
higher order vertices will not contribute as the consequence of the retarded
propagator as well as the charge conservation \cite{Klose:2006dd}. The
resulting Lagrangian has the form

\begin{equation}
\mathcal{L}=\frac{i}{2}(\varphi^{\ast}\partial_{t}\varphi-\partial_{t}
\varphi^{\ast}\varphi)-|\partial_{x}\varphi|^{2}-\frac{g_{1}}{2}
[(\varphi^{\ast}\partial_{x}\varphi)^{2}+(\partial_{x}\varphi^{\ast}
\varphi)^{2}]-\frac{g_{2}}{4}|\varphi|^{2}[(\varphi^{\ast}\partial_{x}
\varphi)^{2}+(\partial_{x}\varphi^{\ast}\varphi)^{2}]-\frac{g_{3}}{2}
|\varphi|^{4}|\partial_{x}\varphi|^{2}  \label{LL3-1}
\end{equation}
where we have introduced, analogous to the two-particle scattering
perturbative calculations, arbitrary coupling constants $g_{1},$ $g_{2},$ $g_{3}$ for each type of vertices to keep track of perturbative calculations.
As in the two-particle scattering case, we should take the limit $g_{i}\rightarrow1$, $i=1,2,3$ in the end. As we will show below, this is a
crucial point, making the S-matrix factorization possible. Clearly, dealing
with more vertices in this case is not an easy task already for the
three-particle case even at the one-loop level. As the number of particles
considered in the scattering increases, one has to expand the action (\ref{LL6}) to higher order, which, as a result, makes the complexity of the
perturbative calculations and the combinatorial analysis practically
impossible. Nevertheless, we make the first step in this direction and show
in the next section the three-particle S-matrix factorization in the first
order of perturbation.

\subsection{Diagrammatic calculations}

\begin{figure}[h]
\centering
\includegraphics[scale=0.65]{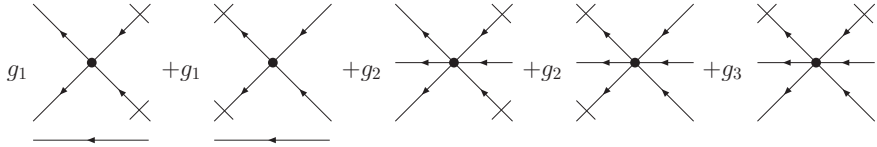}
\caption{The tree-level Feynman diagrams for the three-particle scattering
in the Landau-Lifshitz model.}
\label{treelevelLL3}
\end{figure}

Assuming analyticity of the three-particle scattering amplitude in the 
coupling constants $g_{i}$ it can be written as:
\footnote{In the case of the NLS model, the analyticity of the S-matrix can be seen
from the N-particle coordinate Bethe ansatz calculations of Yang \cite{Yang:1968ev, Yang:1967bm}. In the case of the LL model it is a natural
assumption.}

\begin{equation}
\langle \mathbf{k}|\hat{S}|\mathbf{p}\rangle =\langle \mathbf{k}|\mathbf{p}
\rangle +\langle \mathbf{k}|\hat{S}|\mathbf{p}\rangle \Big|_{g}+\langle 
\mathbf{k}|\hat{S}|\mathbf{p}\rangle \Big|_{g^{2}}+\cdots  \label{LL3-2}
\end{equation}
where $\mathbf{k=\{}k_{1},k_{2},k_{3}\mathbf{\},}$ and $g$ is either of $g_{i}$, $i=1,2,3.$ The non-scattering term has the form: 
\begin{equation}
\langle \mathbf{p}|\mathbf{k}\rangle =3!(2\pi )^{3}\mathcal{S}_{p}[\delta
(p_{1}-k_{1})\delta (p_{2}-k_{2})\delta (p_{3}-k_{3})],  \label{LL3-2a}
\end{equation}
while the tree-level part yields:

\begin{equation}
\langle \mathbf{k}|\hat{S}|\mathbf{p}\rangle \Big|_{g}=9i(2\pi )^{2}\mathcal{S}_{k,p}\left\{ 2g_{1}(k_{1}k_{2}+p_{1}p_{2})2\pi \delta
(k_{3}-p_{3})+g_{2}(k_{1}k_{2}+p_{1}p_{2})-2g_{3}k_{1}p_{1}\right\} \delta
E\delta P  \label{LL3-3}
\end{equation}
We have used here the symmetrization operator defined by \cite{Thacker:1974kv}:

\begin{equation}
\mathcal{S}_{p}\left[ f(\mathbf{p})\right] =\frac{1}{3!}\sum_{A}f(A\mathbf{p}
)  \label{symm}
\end{equation}
where the sum is taken over all possible permutations of $(1,2,3)$, and the
vector $A\mathbf{p}=(p_{A_{1}},p_{A_{2}},p_{A_{3}})$ is the corresponding
set of momenta.\footnote{Without any loss of generality, we choose the momenta $\mathbf{p}$ to be
arranged in the order $p_{1}>p_{2}>p_{3}$.} We have also introduced $\delta
E $ and $\delta P$ to denote the delta functions for the total energy and
momentum conservation, i.e.,

\begin{align}
\delta E& =\delta \left( \sum_{i=1}^{3}(k_{i}^{0}-p_{i}^{0})\right)
\label{LL3-4a} \\
\delta P& =\delta \left( \sum_{i=1}^{3}(k_{i}-p_{i})\right)  \label{LL3-4bb}
\end{align}
The tree level Feynman graphs (connected and disconnected) in the first
order are pictured in Fig. \ref{treelevelLL3}, where the spatial derivatives
are represented by the marks over the particle lines. We stress that each of
these graphs denotes a sum over diagrams of the same topology but with
different permutations of the external momenta.

\begin{figure}[h]
\centering
\includegraphics[scale=0.75]{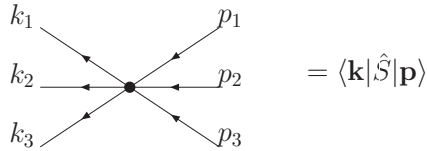}
\caption{Diagrammatic representation of the interaction vertex in momentum
space.}
\label{verticeLL3}
\end{figure}

The expression for the tree level (\ref{LL3-3}) defines the interaction
vertex in the momentum representation, represented by Fig. \ref{verticeLL3}.

\begin{figure}[h]
\centering
\includegraphics[scale=0.75]{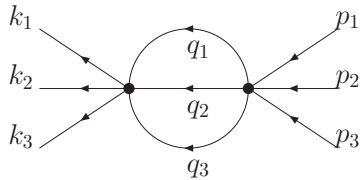}
\caption{Diagrammatic representation of the second order scattering
amplitude.}
\label{2pointLL3}
\end{figure}

Using the interaction vertex in the momentum representation we are able to
compute the second order scattering amplitude as depicted in Fig. \ref{2pointLL3}.

The analytical expression corresponding to the Fig. \ref{2pointLL3} is given
by:

\begin{align}
\langle \mathbf{k}|\hat{S}|\mathbf{p}\rangle \Big|_{g^{2}}& =\frac{1}{6}\int
\prod_{j=1}^{3}\left[ \frac{d^{2}q_{j}}{4\pi ^{2}}\frac{i}{
q_{j}^{0}-q_{j}^{2}+i\epsilon }\right] 9i(2\pi )^{2}\mathcal{S}_{k,q}\left\{
2g_{1}(k_{1}k_{2}+q_{1}q_{2})2\pi \delta
(k_{3}-q_{3})+g_{2}(k_{1}k_{2}+q_{1}q_{2})-\right.  \notag \\
& -\left. 2g_{3}k_{1}q_{1}\right\} \delta \left(
\sum_{i=1}^{3}(k_{i}^{0}-q_{i}^{0})\right) \delta \left(
\sum_{i=1}^{3}(k_{i}-q_{i})\right) \;9i(2\pi )^{2}\mathcal{S}_{q,p}\left\{
2g_{1}(q_{1}q_{2}+p_{1}p_{2})2\pi \delta (q_{3}-p_{3})+\right.  \notag \\
& +\left. g_{2}(k_{1}k_{2}+p_{1}p_{2})-2g_{3}k_{1}p_{1}\right\} \delta
\left( \sum_{i=1}^{3}(q_{i}^{0}-p_{i}^{0})\right) \delta \left(
\sum_{i=1}^{3}(q_{i}-p_{i})\right)  \label{LL3-5}
\end{align}
The delta functions in (\ref{LL3-5}) may be used to integrate over $q_{3}^{0} $ and $q_{3}$, and the integrals over $q_{1}^{0}$ and $q_{2}^{0}$
can be performed by choosing the contour closing in the lower half-plane. We
also symmetrize over $q$, in order to obtain an analytical expression
depending only on the external symmetrized momenta, 

\begin{align}
\langle\mathbf{k}|\hat{S}|\mathbf{p}\rangle\Big|_{g^{2}} & =-\frac{3i}{2}
\delta E\delta P\mathcal{S}_{p,k}\left\{ \int\frac{dq_{1}dq_{2}}{\sum
p^{0}-q_{1}^{2}-q_{2}^{2}-(\sum p-q_{1}-q_{2})^{2}+i\epsilon}\left[ 16\pi
^{2}g_{1}^{2}f_{1}(\mathbf{k},\mathbf{q})f_{1}(\mathbf{p},\mathbf{q})+\right. \right.  \notag \\
& +g_{2}^{2}f_{2}(\mathbf{k},\mathbf{q})f_{2}(\mathbf{p},\mathbf{q}
)+4g_{3}^{2}k_{1}p_{1}\sum k\sum p+4\pi g_{1}g_{2}(f_{1}(\mathbf{k},\mathbf{q})f_{2}(\mathbf{p},\mathbf{q})+f_{2}(\mathbf{k},\mathbf{q})f_{1}(\mathbf{p},\mathbf{q}))-  \notag \\
& -\left. \left. 8\pi g_{1}g_{3}\left( p_{1}\sum pf_{1}(\mathbf{k},\mathbf{q}
)+k_{1}\sum kf_{1}(\mathbf{p},\mathbf{q})\right) -\right. \right.  \notag \\
& -\left. \left. 2g_{2}g_{3}\left( p_{1}\sum pf_{2}(\mathbf{k},\mathbf{q}
)+k_{1}\sum kf_{2}(\mathbf{p},\mathbf{q})\right) \right] \right\}
\label{LL3-6}
\end{align}
where we introduced the following functions:

\begin{align}
f_{1}(\mathbf{x},\mathbf{q}) &
=(x_{1}x_{2}+q_{1}q_{2})\delta(x_{1}+x_{2}-q_{1}-q_{2})+\left[
x_{1}x_{2}+q_{1}\left( \sum x-q_{1}-q_{2}\right) \right] \delta(x_{3}-q_{2})+
\notag \\
& +\left[ x_{1}x_{2}+q_{2}\left( \sum x-q_{1}-q_{2}\right) \right]
\delta(x_{3}-q_{1})  \label{LL3-7a} \\
f_{2}(\mathbf{x},\mathbf{q}) & =3x_{1}x_{2}+(q_{1}+q_{2})\sum
x-q_{1}^{2}-q_{2}^{2}-q_{1}q_{2}  \label{LL3-7b}
\end{align}
To avoid cluttering we do not write the indices in the sum, and use the
notation $\sum x\equiv\underset{i=1}{\overset{3}{\sum}}x_{i}$.

It is easier to compute each term of (\ref{LL3-6}) separately.

\begin{figure}[tbp]
\centering
\includegraphics[scale=0.75]{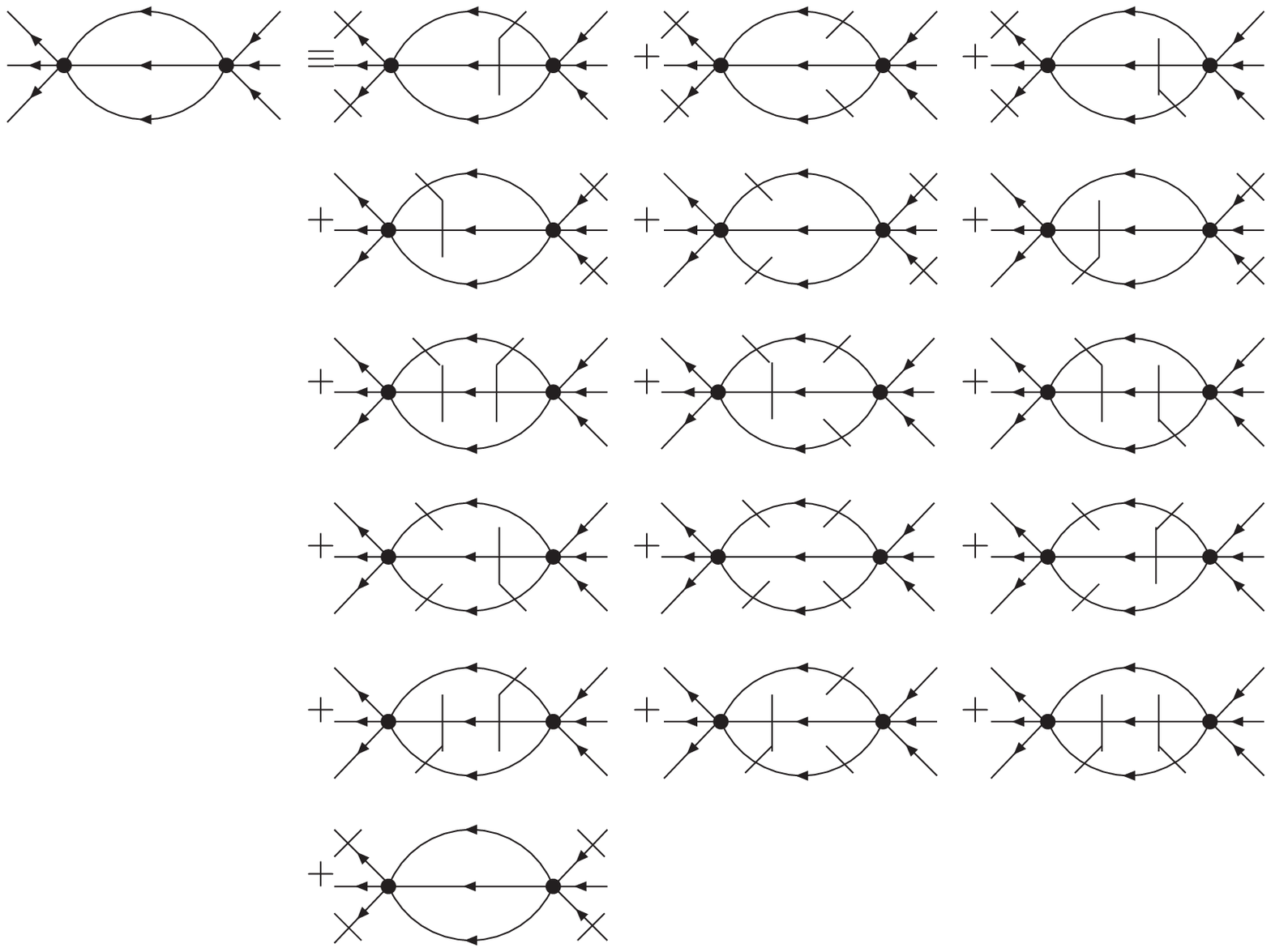}
\caption{The Feynman diagrams proportional to $g_{2}^{2}$, corresponding to
 equation (\protect\ref{LL3-8b}).}
\label{2ndorderapp4}
\end{figure}

\begin{figure}[tbp]
\centering
\includegraphics[scale=0.75]{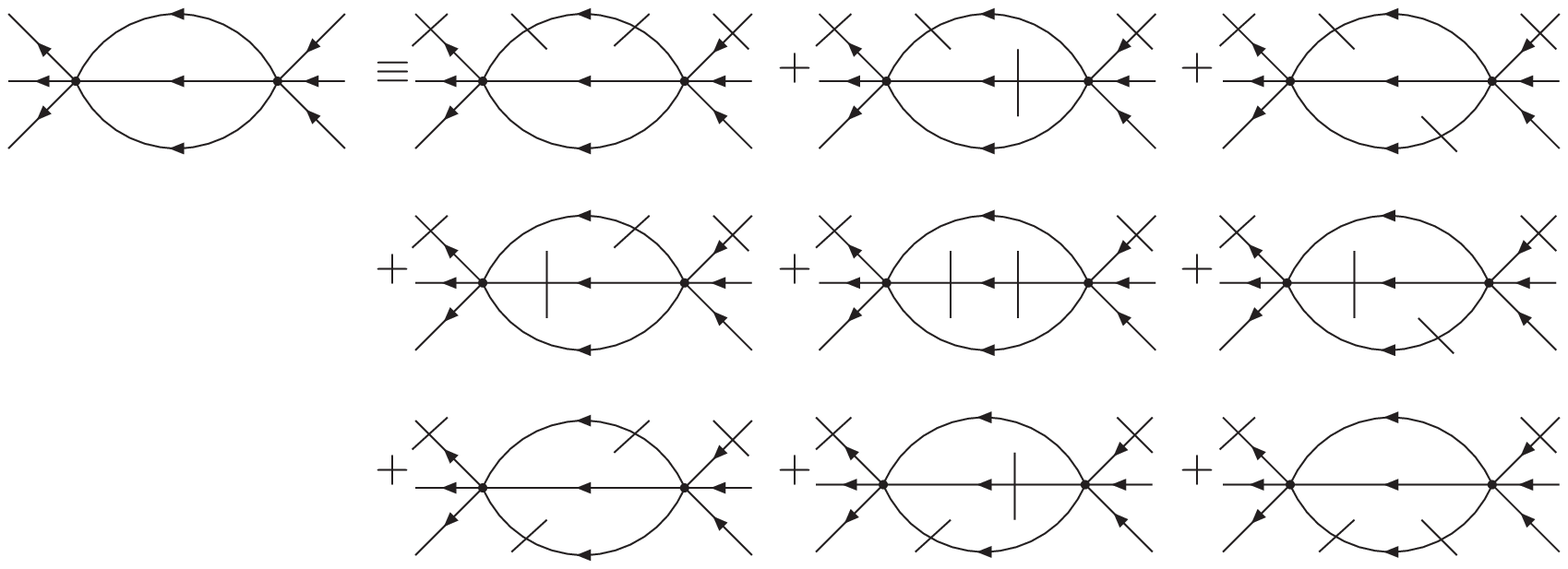}
\caption{The Feynman diagrams proportional to $g_{3}^{2}$, corresponding to
 equation (\protect\ref{LL3-8c}).}
\label{2ndorderapp6}
\end{figure}

\begin{figure}[ptb]
\centering
\includegraphics[scale=0.55]{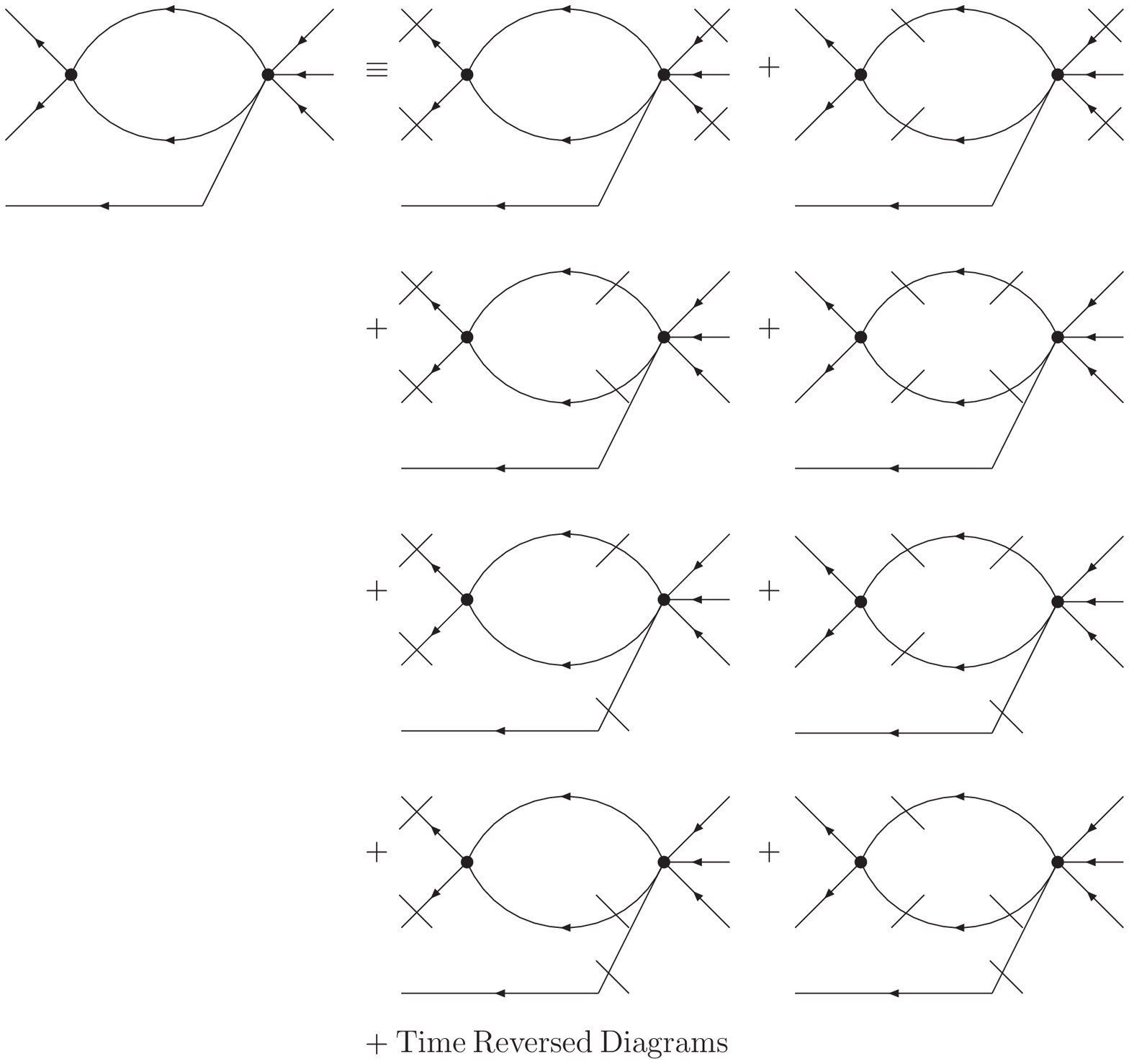}
\caption{The Feynman diagrams proportional to $g_{1} g_{2}$, corresponding
to equation (\protect\ref{LL3-8d}).}
\label{2ndorderapp2}
\end{figure}

\begin{figure}[tbp]
\centering
\includegraphics[scale=0.55]{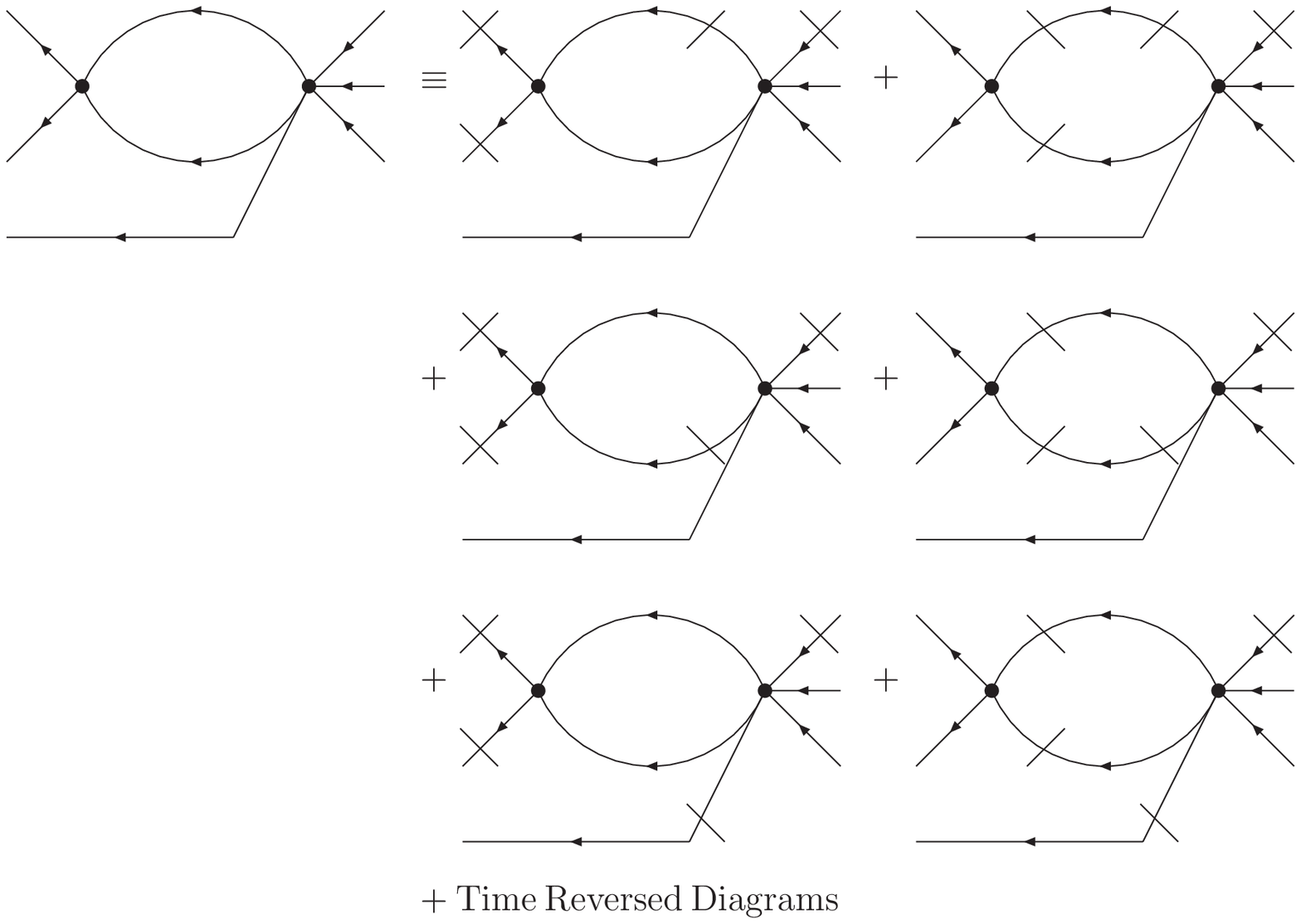}
\caption{The Feynman diagrams proportional to $g_{1}g_{3}$, corresponding to
 equation (\protect\ref{LL3-8e}).}
\label{2ndorderapp3}
\end{figure}

\begin{itemize}
\item[(a)] Term proportional to $g_{1}^{2}$:

Most of the integrals are trivially calculated due to the delta functions in
$f_{1}(\mathbf{x},\mathbf{q})$ , and the remaining delta functions
corresponding to the total energy and momentum conservation allow us to write
these terms as a sum of the following contributions: \emph{i) }the finite
part, which is a function of the external momenta, and \emph{ii) }the term
proportional to $\delta(p_{3}-k_{3})$ . It is the latter term  that may lead
to divergencies, which, however, could be regularized in the same manner as in \cite{Klose:2006dd}. Indeed, we notice that this term can be written in the following form:
\[
-ik_{1}k_{2}p_{1}p_{2}I_{0}(k_{1},k_{2})+i(p_{1}p_{2}+k_{1}k_{2})I_{1}
(k_{1},k_{2})-iI_{2}(k_{1},k_{2}),
\]
where the one-loop integrals $I_{0},$ $I_{1}$ and $I_{2}$ were regularized and
explicitely calculated in \cite{Klose:2006dd}. (We refer the reader to \cite{Klose:2006dd}
for definitions and all calculational details). The finite result after this
regularization has the form (where we keep here the integral representation to be
dealt only in the end of the calculations):
\begin{align}
&  96\pi^{2}g_{1}^{2}\frac{[k_{1}k_{2}+p_{3}(k_{1}+k_{2}-p_{3})][p_{1}
p_{2}+k_{3}(p_{1}+p_{2}-k_{3})]}{\sum p^{0}-p_{3}^{2}-k_{3}^{2}-(p_{1}
+p_{2}-k_{3})^{2}+i\epsilon}-\nonumber\\
&  -24\pi^{2}g_{1}^{2}\int dq\frac{\left[  s-2p_{1}p_{2}\right]  \left[
s+\frac{1}{4}(p_{1}-p_{2})^{2}-\frac{1}{4}(k_{1}-k_{2})^{2}-2k_{1}
k_{2}\right]  }{q^{2}-\frac{1}{4}(p_{1}-p_{2})^{2}-s-i\epsilon}\delta
(k_{3}-p_{3})\label{LL3-8a}
\end{align}

We have denoted $s\equiv\frac{1}{2}\left(  \sum p^{0}-\sum p^{2}\right)  $ and
shifted the integration variable $q\rightarrow q+\nicefrac{1}{2}(k_{1}
+k_{2}).$ The corresponding diagrams are presented in Fig. \ref{2ndorder1}.
Once again, we emphasize that each diagram corresponds to the sum over all
diagrams with the same topology, but with distinct permutations over the
external momenta and different placements of the derivatives. The first
diagram in Fig. \ref{2ndorder1} corresponds to the first term in
(\ref{LL3-8a}), and the second term is the finite part of the second diagram.
The diagrams, corresponding to \ref{2ndorder1}, but with explicit placement of
the derivatives are depicted on Figs. \ref{pv} and \ref{2ndorderapp1}
respectively.\footnote{We draw the remaining Feynman diagrams in the
\textbf{Appendix}.}

\begin{figure}[tbp]
\centering
\includegraphics[scale=0.65]{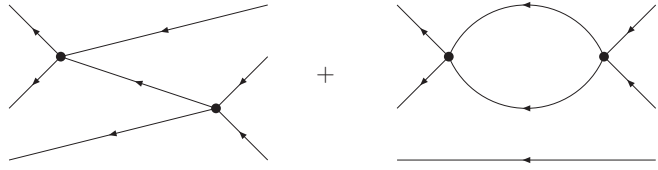}
\caption{Feynman diagram corresponding to the equation (\protect\ref{LL3-8a}).}
\label{2ndorder1}
\end{figure}

\item[(b)] Term proportional to $g_{2}^{2}$:

After some algebraic manipulations this term can be written in the form:

\begin{align}
& g_{2}^{2}\int \frac{f_{2}(\mathbf{k},\mathbf{q})f_{2}(\mathbf{p},\mathbf{q})\;dq_{1}dq_{2}}{\sum p^{0}-q_{1}^{2}-q_{2}^{2}-(\sum
p-q_{1}-q_{2})^{2}+i\epsilon }  \notag \\
& =-\frac{g_{2}^{2}}{2}\int dq_{1}dq_{2}\frac{\left[ 3p_{1}p_{2}-s\right] 
\left[ 3k_{1}k_{2}-s\right] }{q_{1}^{2}+q_{2}^{2}+q_{1}q_{2}-(q_{1}+q_{2})
\sum p-s-i\epsilon }  \label{LL3-8b}
\end{align}
The corresponding Feynman diagram is represented by Fig. \ref{2ndorderapp4}.

\item[(c)] Term proportional to $g_{3}^{2}$:

Simple transformations lead to the following expression:

\begin{align}
4& g_{3}^{2}\int \frac{k_{1}p_{1}\sum k\sum p\;dq_{1}dq_{2}}{\sum
p^{0}-q_{1}^{2}-q_{2}^{2}-(\sum p-q_{1}-q_{2})^{2}+i\epsilon }  \notag \\
& =-2g_{3}^{2}\int \frac{k_{1}p_{1}\sum k\sum p\;dq_{1}dq_{2}}{q_{1}^{2}+q_{2}^{2}+q_{1}q_{2}-(q_{1}+q_{2})\sum p-s-i\epsilon }
\label{LL3-8c}
\end{align}
The corresponding Feynman diagram is represented by Fig. \ref{2ndorderapp6}.

\item[(d)] Term proportional to $g_{1}g_{2}$:

One of the integrals in this term is trivially computed using the delta
functions of $f_{1}(\mathbf{x},\mathbf{q})$, while the remaining integral
after some algebraic manipulations and the shift of the integration variable $q_1\rightarrow q_1 + \nicefrac{1}{2}(k_{1}+k_{2})$ and $q_2\rightarrow q_2 + \nicefrac{1}{2}(p_{1}+p_{2})$, yields:

\begin{align}
& 4\pi g_{1}g_{2}\int dq_{1}dq_{2}\;\frac{f_{1}(\mathbf{k},\mathbf{q})f_{2}(\mathbf{p},\mathbf{q})+f_{2}(\mathbf{k},\mathbf{q})f_{1}(\mathbf{p},\mathbf{q})}{\sum p^{0}-q_{1}^{2}-q_{2}^{2}-(\sum p-q_{1}-q_{2})^{2}+i\epsilon } 
\notag \\
=& -6\pi g_{1}g_{2}\int dq\;\frac{\left[ k_{3}(k_{1}+k_{2})-k_{1}k_{2}+s\right] \left[ s-3p_{1}p_{2}\right] }{q^{2}-\frac{1}{4}(k_{1}+k_{2})^{2}-k_{3}(k_{1}+k_{2})-s-i\epsilon }-  \notag \\
& -6\pi g_{1}g_{2}\int dq\;\frac{\left[ p_{3}(p_{1}+p_{2})-p_{1}p_{2}+s\right] \left[ s-3k_{1}k_{2}\right] }{q^{2}-\frac{1}{4}
(p_{1}+p_{2})^{2}-p_{3}(p_{1}+p_{2})-s-i\epsilon }  \label{LL3-8d}
\end{align}

The corresponding Feynman diagram is represented by Fig. \ref{2ndorderapp2}.

\item[(e)] Term proportional to $g_{1}g_{3}$:

Once again the delta functions in $f_{1}(\mathbf{x},\mathbf{q})$ allow us
to trivially compute one of the integrals and the resulting expression has
the form:

\begin{align}
& -8\pi g_{1}g_{3}\int dq_{1}dq_{2}\;\frac{p_{1}\sum pf_{1}(\mathbf{k},\mathbf{q})+k_{1}\sum kf_{1}(\mathbf{p},\mathbf{q})}{\sum
p^{0}-q_{1}^{2}-q_{2}^{2}-(\sum p-q_{1}-q_{2})^{2}+i\epsilon }  \notag \\
& =-12\pi g_{1}g_{3}\int dq\left\{ p_{1}\sum p+k_{1}\sum k\right\} -  \notag
\\
& -12\pi g_{1}g_{3}\int dq\frac{k_{1}\sum k\left[
p_{3}(p_{1}+p_{2})-p_{1}p_{2}+s\right] }{q^{2}-\frac{1}{4}
(p_{1}+p_{2})^{2}-p_{3}(p_{1}+p_{2})-s-i\epsilon }  \notag \\
& -12\pi g_{1}g_{3}\int dq\frac{p_{1}\sum p\left[
k_{3}(k_{1}+k_{2})-k_{1}k_{2}+s\right] }{q^{2}-\frac{1}{4}
(k_{1}+k_{2})^{2}-k_{3}(k_{1}+k_{2})-s-i\epsilon }  \label{LL3-8e}
\end{align}

The corresponding Feynman diagram is represented by Fig. \ref{2ndorderapp3}.

\item[(f)] Term proportional to $g_{2}g_{3}$:

There are no delta functions present in this term and after some
transformations we write it in the form:

\begin{align}
& -2g_{2}g_{3}\int \frac{p_{1}\sum pf_{2}(\mathbf{k},\mathbf{q})+k_{1}\sum
kf_{2}(\mathbf{p},\mathbf{q})}{\sum p^{0}-q_{1}^{2}-q_{2}^{2}-(\sum
p-q_{1}-q_{2})^{2}+i\epsilon }dq_{1}dq_{2} \nonumber \\ 
& =-g_{2}g_{3}\int dq_{1}dq_{2}\frac{\left( p_{1}\sum p+k_{1}\sum k\right)
s-3k_{1}k_{2}p_{1}\sum p-3k_{1}p_{1}p_{2}\sum k}{q_{1}^{2}+q_{2}^{2}+q_{1}q_{2}-(q_{1}+q_{2})\sum p-s-i\epsilon }\label{LL3-8f} 
\end{align}

The corresponding Feynman diagram represented by Fig. \ref{2ndorderapp5}.
\end{itemize}

Let us note, that in equations (\ref{LL3-8a}), (\ref{LL3-8b}), (\ref{LL3-8d}), (\ref{LL3-8e}) and (\ref{LL3-8f}) we have ignored the divergences and
kept only the finite parts. Although the above formulas have been obtained
for the general off-shell case, putting the external momenta on shell, i.e., 
$p_{i}^{0}=p_{i}^{2}$ and $k_{i}^{0}=k_{i}^{2}$, for $i=1,2,3$ leads to
considerable simplifications. 
This is due to the following identities:

\begin{align}
\sum p^{0}-\sum p^{2}=\sum k^{0}-\sum k^{2} & =0  \notag \\
\sum p-\left( \sum p\right) ^{2}=-2(p_{1}p_{2}+p_{1}p_{3}+p_{2}p_{3}) &
,\sum k-\left( \sum k\right) ^{2}=-2(k_{1}k_{2}+k_{1}k_{3}+k_{2}k_{3})
\label{LL3-9} \\
\sum p-\left( \sum p\right) ^{2}=\sum k-\left( \sum k\right) ^{2} &
\Rightarrow p_{1}p_{2}+p_{1}p_{3}+p_{2}p_{3}=k_{1}k_{2}+k_{1}k_{3}+k_{2}k_{3}
\notag
\end{align}
The remaining integrals are not difficult to compute. In fact, the integral
that one has to compute in the second term of (\ref{LL3-8a}) is simply,

\begin{equation}
\int\frac{dq}{q^{2}-\frac{1}{4}(p_{1}-p_{2})^{2}-i\epsilon}=\frac{2\pi i}{|p_{1}-p_{2}|}  \label{LL3-10a}
\end{equation}
The remaining integrals in (\ref{LL3-8d}) and (\ref{LL3-8e}) are:

\begin{align}
\int\frac{dq}{q^{2}-\frac{1}{4}(k_{1}+k_{2})^{2}-k_{3}(k_{1}+k_{2})+k_{1}k_{2}+k_{1}k_{3}+k_{2}k_{3}-i
\epsilon} & =\frac{2\pi i}{|k_{1}-k_{2}|}  \label{LL3-10b} \\
\int\frac{dq}{q^{2}-\frac{1}{4}
(p_{1}+p_{2})^{2}-p_{3}(p_{1}+p_{2})+p_{1}p_{2}+p_{1}p_{3}+p_{2}p_{3}-i\epsilon} & =\frac{2\pi i}{|p_{1}-p_{2}|}  \label{LL3-10b1}
\end{align}
Finally the remaining double integral in (\ref{LL3-8b}), (\ref{LL3-8c}) e (\ref{LL3-8f}) is:

\begin{equation}
\int \frac{dq_{1}dq_{2}}{q_{1}^{2}+q_{2}^{2}+q_{1}q_{2}-p(q_{1}+q_{2})+(p_{1}p_{2}+p_{1}p_{3}+p_{2}p_{3})-i\epsilon 
}=-\frac{4\pi }{\sqrt{3}}\left[ \frac{i\pi }{2}-\frac{1}{2}\log Q^{2}+\frac{1}{2}\lim_{R\rightarrow \infty }\log (R^{2}-Q^{2})\right]  \label{LL3-10c}
\end{equation}
where we denoted:

\begin{equation}
Q^{2}=\frac{4}{6}\left[
(p_{1}-p_{2})^{2}+(p_{1}-p_{3})^{3}+(p_{2}-p_{3})^{2} \right]  \label{LL3-11}
\end{equation}
Keeping only the finite part in the integral (\ref{LL3-10c}), and collecting
the above expressions, we find that the second order scattering amplitude
has the form:

\begin{align}
\langle\mathbf{k}| \hat{S} | \mathbf{p} \rangle\Big|_{g^{2}} = - \frac{3}{2} 
\mathcal{S}_{p,k} \left\{ 96 \pi^{2} g_{1}^{2} \left[ \frac{[k_{1}k_{2} +
p_{3}(k_{1}+k_{2}-p_{3})][p_{1}p_{2}+ k_{3}(p_{1}+p_{2}-k_{3})]}{p_{1}^{2} +
p_{2}^{2} - k_{3}^{2} - (p_{1}+p_{2}-k_{3})^{2} + i \epsilon} - \frac {k_{1}k_{2}p_{1}p_{2}}{p_{1}-p_{2}} 2 \pi i \delta(p_{3}-k_{3}) \right]
\right. +  \notag \\
+ \left. k_{1}p_{1} \left[ 48 \pi^{2} i g_{1}g_{3} \left( \frac{k_{1}p_{2}}{p_{1}-p_{2}} + \frac{k_{2}p_{1}}{|k_{1}-k_{2}|} \right) + \frac{4\pi }{\sqrt{3}} (i \pi- \log Q^{2}) \left(
(k_{1}p_{2}+k_{2}p_{1})g_{3}(2g_{3}-3g_{2}) + g_{3}^{2} k_{1}p_{1} \right) \right] \right. +  \notag \\
+ \left. k_{1}k_{2}p_{1}p_{2} \left[ 48 \pi^{2} i \left( \frac{1}{p_{1}-p_{2}} + \frac{1}{|k_{1}-k_{2}|}\right) g_{1}(2g_{3} - 3g_{2}) + 
\frac{4\pi}{\sqrt{3}}(i \pi- \log Q^{2})(2g_{3} - 3g_{2})^{2} \right]
\right\} \delta E \delta P  \label{LL3-12}
\end{align}

\subsection{The S-matrix factorization}

Had we assumed the quantum integrability of the LL\ model, we could have
obtained the S-matrix for three particle scattering by using the relation:

\begin{equation}
\langle \mathbf{k}|\hat{S}|\mathbf{p}\rangle =S(\mathbf{p})\delta _{+}^{(3)}(\mathbf{p},\mathbf{k})  \label{LL3-13}
\end{equation}
where $\delta _{+}^{(3)}(\mathbf{p},\mathbf{k})=\langle \mathbf{p}|\mathbf{k}
\rangle =3!(2\pi )^{3}\mathcal{S}_{p}[\delta (p_{1}-k_{1})\delta
(p_{2}-k_{2})\delta (p_{3}-k_{3})]$. This relation, analogous to (\ref{LL12}), underlines the fact that in the scattering process the particle
annihilation and creation are not possible, and the set of momenta before
and after the scattering is the same. This is not obvious from the
calculations presented in the previous section. Indeed, although the
non-scattering term clearly has the form (\ref{LL3-13}), neither the tree
level (\ref{LL3-3}), nor the second order (\ref{LL3-12}) terms are
manifestly of this form. Nevertheless, we will show that this is indeed the
case, and one can reduce the expressions for the tree level (\ref{LL3-3})
and the second order (\ref{LL3-12}) contributions to the forms that are
proportional to $\delta _{+}^{(3)}(\mathbf{p},\mathbf{k})$.

Let us first focus on the tree level term (\ref{LL3-3}). Expanding
explicitly the symmetrization operator $\mathcal{S}_{k,p},$ we find:

\begin{align}
\langle \mathbf{k}|\hat{S}|\mathbf{p}\rangle \Big|_{g}& =2ig_{1}\left[ \frac{p_{1}p_{2}}{p_{1}-p_{2}}+\frac{p_{1}p_{3}}{p_{1}-p_{3}}+\frac{p_{2}p_{3}}{p_{2}-p_{3}}\right] \delta _{+}^{(3)}(\mathbf{p},\mathbf{k})+  \notag \\
& +2i(2\pi )^{2}\left\{
(p_{1}p_{2}+p_{1}p_{3}+p_{2}p_{3})(3g_{2}-2g_{3})-g_{3}(p_{1}^{2}+p_{2}^{2}+p_{3}^{2})\right\} \delta E\delta P
\label{LL3-14}
\end{align}
We immediately notice that the two terms in above expression are quite
different. Namely, the first term of (\ref{LL3-14}) is already in the form
needed for the S-matrix to satisfy (\ref{LL3-13}), while the second term, on the other hand, does not satisfy this condition.
Therefore, had we restricted ourselves only to this order, the condition (\ref{LL3-13}), necessary for integrability to hold, would have failed.
Similarly, it is not difficult to see that the second order scattering
amplitude also contains terms that are not proportional to $\delta
_{+}^{(3)}(\mathbf{p},\mathbf{k})$, such as the terms in the second and
third square brackets in (\ref{LL3-12}).

Let us note, however, that if the quantum integrability, and as a
consequence, the S-matrix factorization were true, the three-particle
S-matrix would have the exact form:

\begin{align}
S^{(3)}(\mathbf{p},\mathbf{k})&
=S^{(2)}(p_{1},p_{2})S^{(2)}(p_{1},p_{3})S^{(2)}(p_{2},p_{3})  \notag \\
& =\left( \frac{\frac{1}{p_{1}}-\frac{1}{p_{2}}-ig}{\frac{1}{p_{1}}-\frac{1}{p_{2}}+ig}\right) \left( \frac{\frac{1}{p_{1}}-\frac{1}{p_{3}}-ig}{\frac{1}{p_{1}}-\frac{1}{p_{3}}+ig}\right) \left( \frac{\frac{1}{p_{2}}-\frac{1}{p_{3} }-ig}{\frac{1}{p_{2}}-\frac{1}{p_{3}}+ig}\right)  \notag \\
& = 1+2\sum_{n=1}^{2}\left[ ig\left( \frac{p_{1}p_{2}}{p_{1}-p_{2}}+\frac{p_{1}p_{3}}{p_{1}-p_{3}}+\frac{p_{2}p_{3}}{p_{2}-p_{3}}\right) \right]
^{n}+O(g^{3})  \label{LL3-15}
\end{align}
where we have expanded in the last line the exact expression up to the
second order in $g$. Comparing the equations (\ref{LL3-14}) and (\ref{LL3-15}
), we see that the unity in (\ref{LL3-15}) corresponds to the non-scattering
part $\langle \mathbf{p}|\mathbf{k}\rangle ,$ while the first term in (\ref{LL3-14}) corresponds to the first order term of (\ref{LL3-15}). We will now
demonstrate, that the second term in (\ref{LL3-14}) is such that it can be
canceled out exactly with a term from the second order contribution (\ref{LL3-12}).

\begin{figure}[tbp]
\centering
\includegraphics[scale=0.55]{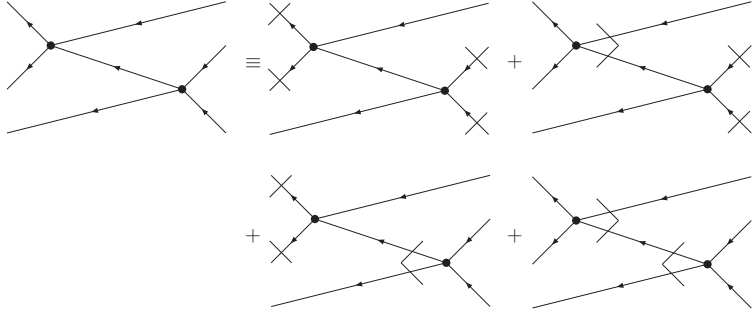}
\caption{The Feynman diagrams for the first term in the equation (\protect
\ref{LL3-12}) with all derivatives placed.}
\label{pv}
\end{figure}

A very similar expression for first term of (\ref{LL3-12}) appears in the
NLS model (see formula (13) of \cite{Thacker:1974kv}). In the latter case,
the use of the formula 
\begin{equation}
\frac{1}{x\pm i0}=\mp i\pi \delta (x)+\mathrm{P.V.}\left( \frac{1}{x}\right)
\label{sochotski}
\end{equation}
proved to be crucial to show the factorization (see also \cite{Puletti:2007hq}). We, thus, expect that the usage of this formula should
play the same role in our case. However, while the principal value part has
no contribution for the NLS\ model, in the LL\ model it plays a non-trivial
role and leads to the cancellations which make the factorization possible.
We draw all the Feynman graphs for this term in Fig. \ref{pv}, where only
the sum over topologically similar diagrams with different permutations of
the external momenta is implicit. Thus, there are four diagrams, which
correspond to all possible placements of the derivatives.

We remind that we have chosen the momenta $\mathbf{p}$ to be arranged in
the order $p_{1}>p_{2}>p_{3}$ for the three-particle scattering to be
possible. The principal value of the first term in (\ref{LL3-12}) can be
computed if we take $k_{i}\neq p_{j}$, $i,j\in \{1,2,3\}$ and use the
identity 
\begin{equation*}
\frac{i}{k_{i}^{2}+k_{j}^{2}-p_{l}^{2}-(k_{i}+k_{j}-p_{l})^{2}+i\epsilon }= 
\frac{1}{2}\frac{i}{k_{i}-k_{j}}\left( \frac{1}{p_{l}-k_{j}-i\epsilon }+ 
\frac{1}{k_{i}-p_{l}-i\epsilon }\right) ,
\end{equation*}
as well as take into account the delta functions corresponding to the energy
and momentum conservation. Only the first and last diagrams on the R.H.S
of the Fig. \ref{pv} will contribute to the principal value, while the other
two vanish identically due to the momentum conservation. After some tedious
transformations one finds:

\begin{equation}
\mathrm{P.V.}\left\{ \mathcal{S}_{p,k}\left[ \frac{[k_{1}k_{2}+p_{3}(k_{1}+k_{2}-p_{3})][p_{1}p_{2}+k_{3}(p_{1}+p_{2}-k_{3})]}{p_{1}^{2}+p_{2}^{2}-k_{3}^{2}-(p_{1}+p_{2}-k_{3})^{2}+i\epsilon}\right]
\right\} =\frac {1}{18}
(-p_{1}^{2}-p_{2}^{2}-p_{3}^{2}+p_{1}p_{2}+p_{1}p_{3}+p_{2}p_{3})
\label{LL3-16}
\end{equation}
The delta function part of (\ref{sochotski}) yields:

\begin{equation}
\delta \left( p_{1}^{2}+p_{2}^{2}-k_{3}^{2}-(p_{1}+p_{2}-k_{3})\right) =\frac{1}{2(p_{1}-p_{2})}\left[ \delta (p_{1}-k_{3})+\delta (p_{2}-k_{3}) 
\right]  \label{LL3-17}
\end{equation}
Substituting this result back into (\ref{LL3-12}) we find:

\begin{align}
\langle \mathbf{k}|\hat{S}|\mathbf{p}\rangle \Big|_{g^{2}}& =8i\pi
^{2}g_{1}^{2}(p_{1}^{2}+p_{2}^{2}+p_{3}^{2}-p_{1}p_{2}-p_{1}p_{3}-p_{2}p_{3})\delta E\delta P-
\notag \\
& -\frac{3i}{2}\delta E\delta P\;\mathcal{S}_{p,k}\left\{ -96\pi
^{2}g_{1}^{2}\frac{k_{1}k_{2}p_{1}p_{2}}{p_{1}-p_{2}}2\pi i\left[ \delta
(k_{3}-p_{1})+\delta (k_{3}-p_{2})+\delta (k_{3}-p_{3})\right] \right. 
\notag \\
& +\left. k_{1}p_{1}\left[ 48\pi ^{2}ig_{1}g_{3}\left( \frac{k_{1}p_{2}}{
p_{1}-p_{2}}+\frac{k_{2}p_{1}}{|k_{1}-k_{2}|}\right) +\frac{4\pi }{\sqrt{3}}
(i\pi -\log Q^{2})\left(
(k_{1}p_{2}+k_{2}p_{1})g_{3}(2g_{3}-3g_{2})+g_{3}^{2}k_{1}p_{1}\right) \right] \right. +  \notag \\
& +\left. k_{1}k_{2}p_{1}p_{2}\left[ 48\pi ^{2}i\left( \frac{1}{p_{1}-p_{2}}
+ \frac{1}{|k_{1}-k_{2}|}\right) g_{1}(2g_{3}-3g_{2})+\frac{4\pi }{\sqrt{3}}
(i\pi -\log Q^{2})(2g_{3}-3g_{2})^{2}\right] \right\}  \label{LL3-18}
\end{align}
To proceed, we explicitly expand the symmetrization operator $\mathcal{S}_{k,p}$ and obtain the following form for the second order scattering
amplitude :

\begin{align}
\langle \mathbf{k}|\hat{S}|\mathbf{p}\rangle \Big|_{g^{2}}& =8i\pi
^{2}g_{1}^{2}(p_{1}^{2}+p_{2}^{2}+p_{3}^{2}-p_{1}p_{2}-p_{1}p_{3}-p_{2}p_{3})\delta E\delta P+
\notag \\
& +2i^{2}g_{1}^{2}\left[ \frac{p_{1}p_{2}}{p_{1}-p_{2}}+\frac{p_{1}p_{3}}{p_{1}-p_{3}}+\frac{p_{2}p_{3}}{p_{2}-p_{3}}\right] ^{2}\delta _{+}^{(3)}(\mathbf{p},\mathbf{k})-\frac{3i}{2}\mathcal{R}(g^{2})\delta E\delta P
\label{LL3-19}
\end{align}
where we have separated the one loop contribution:

\begin{align}
& \mathcal{R}(g^{2})  \notag \\
& =-\frac{16}{3}\pi ^{2}ig_{1}\left\{ \left[ \frac{p_{1}p_{2}}{p_{1}-p_{2}}+ 
\frac{p_{1}p_{3}}{p_{1}-p_{3}}+\frac{p_{2}p_{3}}{p_{2}-p_{3}}\right] \left[
(3g_{2}-2g_{3})(k_{1}k_{2}+k_{1}k_{3}+k_{2}k_{3})-g_{3}(k_{1}^{2}+k_{2}^{2}+k_{3}^{2}) \right] \right. +  \notag \\
& +\left. \left[ \frac{k_{1}k_{2}}{|k_{1}-k_{2}|}+\frac{k_{1}k_{3}}{
|k_{1}-k_{3}|}+\frac{k_{2}k_{3}}{|k_{2}-k_{3}|}\right] \left[
(3g_{2}-2g_{3})(p_{1}p_{2}+p_{1}p_{3}+p_{2}p_{3})-g_{3}(p_{1}^{2}+p_{2}^{2}+p_{3}^{2}) \right] \right\} +  \notag \\
& +\frac{4\pi }{\sqrt{3}}(i\pi -\log Q^{2})\left[
(3g_{2}-2g_{3})(p_{1}p_{2}+p_{1}p_{3}+p_{2}p_{3})-g_{3}(p_{1}^{2}+p_{2}^{2}+p_{3}^{2}) \right] \times  \notag \\
& \times \left[
(3g_{2}-2g_{3})(k_{1}k_{2}+k_{1}k_{3}+k_{2}k_{3})-g_{3}(k_{1}^{2}+k_{2}^{2}+k_{3}^{2}) \right]  \label{LL3-20}
\end{align}
We note that in the limit $g_{1}=g_{2}=g_{3}=1$ the first term of (\ref{LL3-19}) has the opposite sign of the second term in (\ref{LL3-14}). Thus,
the unwanted term indeed cancels out and the remaining terms combine into the
form:

\begin{equation}
\langle \mathbf{k}|\hat{S}|\mathbf{p}\rangle =\left[ 1+2\sum_{n=1}^{2}\left[
i\left( \frac{p_{1}p_{2}}{p_{1}-p_{2}}+\frac{p_{1}p_{3}}{p_{1}-p_{3}}+\frac{p_{2}p_{3}}{p_{2}-p_{3}}\right) \right] ^{n}\right] \delta _{+}^{(3)}(
\mathbf{p},\mathbf{k})-\frac{3i}{2}\mathcal{R}(g^{2})\delta E\delta P+O(g^{3})  \label{LL3-21}
\end{equation}
The first term in (\ref{LL3-21}) is the three-particle scattering S-matrix
in the second order, consistent with the factorization relation (\ref{LL3-15}). Thus, we conclude that to have the S-matrix factorization already in the
lowest order, the terms from the higher order contribution should be taken
into account to cancel out the unwanted terms. Diagrammatically, by making
use of the relation (\ref{sochotski}) we have found that the principal value
of the Feynman diagrams depicted in Fig. \ref{pv} cancels out the terms at
first order that prevented the factorizability of the S-matrix at this
order, while the contribution of the delta-function term of (\ref{sochotski}
), added to the term in Fig. \ref{2ndorderapp1}, gave the factorizable
second order contribution to the S-matrix. Although we left out the $\mathcal{R}(g^{2})$ contribution in (\ref{LL3-21}), this calculation
provides us with a remarkable scheme in which the contributions of different
orders cancel each other to yield S-matrix factorizability.

It is not difficult to see that the diagrams that we had to sum up to
obtain the three-particle S-matrix in the first order are of the order $\hbar^{0}$ since neither the tree level, nor the diagrams in Fig. \ref{pv}
contain any loops. Clearly, the cancellation at higher orders may happen for
the diagrams of the same order in $\hbar.$ Even though we were able to prove
the factorizability at the first order, the complexity of diagrammatic and
combinatorial analysis make the computations of higher orders practically
impossible to carry out, and another approach has to be employed. This
subject is currently under investigation.

\section{Conclusion and discussion}

\label{conclusion}

In this paper we have shown the three-particle S-matrix factorization at the
first order for the LL model. The calculations we have presented show the
non-trivial mechanism behind the factorization even in the lowest order of
perturbation. In the process we have also shown the absence of particle
annihilation and creation in the scattering processes, and that the set of
momenta before and after the scattering, correspondingly, is unchanged. Our
consideration was motivated by the difficulties to analyze the quantum
integrability of the LL model in the framework of the quantum inverse
scattering method. The highly singular nature of the LL model does not allow
the use of the trace identities and, as a result, the set of local operators
in involution is in general hard to derive. In fact, only the action of the
quantum Hamiltonian on the two-particle sector is known. In addition, the
standard two-particle state (\ref{2pwf}), underlining the clear meaning of
the scattering process, is not valid for the LL model quantized in terms of
the fields $\varphi(x)$ (\ref{LL5}), namely, the particles created by the
operator $\varphi ^{+}(x)$ are not the Bethe particles. As the S-matrix
factorization, which expresses quantum integrability, relies on the presence
of an infinite set of commuting local operators, it is important to
independently verify quantum integrability using field theoretic
methods. Let us note, that unlike the LL model, these subtleties are absent in the
classically equivalent NLS model, and the quantum inverse
scattering method is easily shown to be consistent with the field theoretic
calculations.

The next obvious problem is generalizing the above construction to all
orders of perturbation, as well as showing the factorization for the $N$-particle scattering amplitude. This is not an easy task, as even in the
lowest order the difficulties of diagrammatic and combinatorial nature are
quite complex. The essential difference with the NLS model is the new
vertices that appear in each order for the LL model. As we have shown, the
factorization mechanism is such that the higher order terms cancel the
unwanted terms via the use of the formula (\ref{sochotski}). Clearly, at
higher orders the analysis will be too complicated to be carried out.
Thus, an alternative framework is needed to perform the diagrammatic
calculations. The latter have to be also carried out to understand the
effects of renormalization. The reason is twofold. First, as we have shown,
the coupling constants $g_i, i=1,2,3$, which we had introduced by hand to
keep track of the perturbative order, had to be set to unity for the
three-particle S-matrix to be factorizable. Although we have done so without
properly taking care of the renormalization, it is important to carry out
all the necessary calculations to show the validity of this assumption.
Secondly, it is clear that the singularities arising in the quantum inverse
scattering method, when deriving the local commuting operators, should
exhibit themselves in the renormalization of the coupling constants and
fields. Thus, it would be interesting to establish a direct map between the
two methods. We stress the importance of this program in the context of the
AdS/CFT correspondence, since the difficulties present in the quantization
of LL model will also be present in the higher sectors of the $AdS_{5}\times
S^{5}$ string.

\section*{Acknowledgments}

One of us (A.M.) would like to thank Prof. Ashok Das for many useful
discussions. The work of A.M. was supported by the FAPESP grant No.
05/05147-3. The work of A.P. was supported by the FAPESP grant No.
06/56056-0. The work of V.O.R. is supported by CNPq, FAPESP and PROSUL grant
No. 490134/2006-8. The work of G.W. was supported by the FAPESP grant No.
06/02939-9.





\section*{Appendix: Feynman diagrams}

\label{appendix} \addcontentsline{toc}{section}{Appendix: Feynman diagrams}

\begin{figure}[h]
\centering
\includegraphics[scale=0.55]{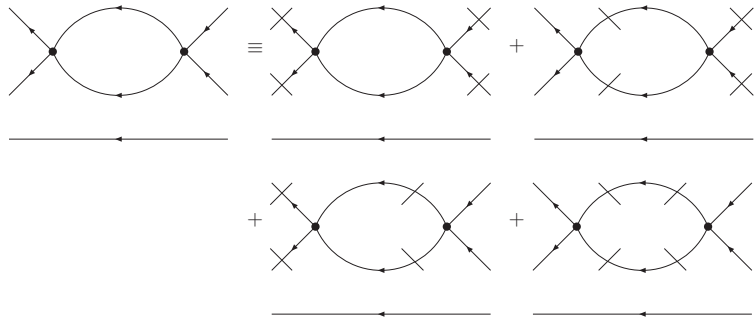}
\caption{The Feynman diagrams with the correct placement of derivatives 
corresponding to the second Feynman graph pictured on Fig. \protect\ref{2ndorder1}.}
\label{2ndorderapp1}
\end{figure}

In this appendix we draw the remaining Feynman graphs for the three-particle
second order scattering amplitude with the explicit placement of
derivatives. We note, however, that the sum over the diagrams of the same
topology, but with different permutations of the external momenta, is
assumed. The two diagrams depicted in Fig. \ref{2ndorder1}, which correspond, respectively, to each of the finite terms of the equation (\ref{LL3-8a}), are represented, with all derivatives placed, by Figs. \ref{pv}
and \ref{2ndorderapp1}. 

The following list of figures contains the complete Feynman diagrams
corresponding to the equations (\ref{LL3-8b}-\ref{LL3-8f}).

\begin{figure}[tbp]
\centering
\includegraphics[scale=0.75]{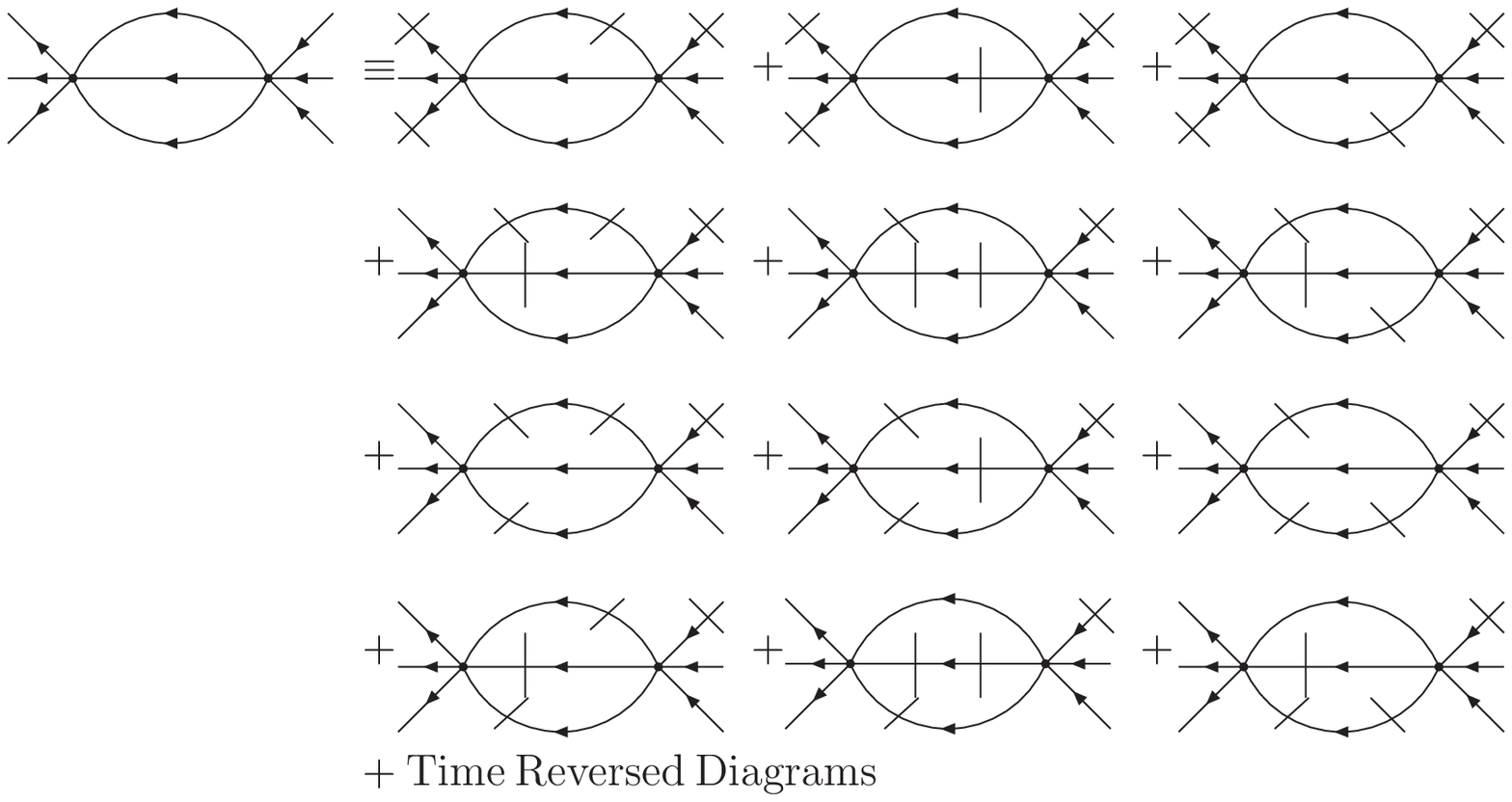}
\caption{The Feynman diagrams proportional to $g_{2}g_{3}$, corresponding to
 equation (\protect\ref{LL3-8f}).}
\label{2ndorderapp5}
\end{figure}

\clearpage
\bibliographystyle{JHEP}
\bibliography{ll_factorization_published}
\end{document}